\documentclass[twocolumn,aps,prl,superscriptaddress,longbibliography]{revtex4-1}
\usepackage{verbatim}
\usepackage{color}
\usepackage{amsmath,amssymb,bm,braket,float,mathtools,subfigure}
\usepackage{graphicx,xfrac,appendix}
\usepackage[english]{babel}
\usepackage{physics}

\makeatletter

\usepackage{epstopdf,float}

\makeatother
\usepackage[colorlinks,bookmarks=false,citecolor=blue,linkcolor=red,urlcolor=blue]{hyperref}

\begin{document}
\title{Quantum Mpemba effect in quasiperiodic systems}
\author{Ao Zhou}
\affiliation{Department of Physics, Zhejiang Normal University, Jinhua 321004, China}
\author{Feng Lu}
\affiliation{Department of Physics, Zhejiang Normal University, Jinhua 321004, China}

\author{Shujie Cheng}
\thanks{chengsj@zjnu.edu.cn}
\affiliation{Xingzhi College, Zhejiang Normal University, Lanxi 321100, China}
\affiliation{Department of Physics, Zhejiang Normal University, Jinhua 321004, China}

\author{Gao Xianlong}
\thanks{gaoxl@zjnu.edu.cn}
\affiliation{Department of Physics, Zhejiang Normal University, Jinhua 321004, China}

\begin{abstract}
We study a one-dimensional quasiperiodic tight-binding model with simultaneous off-diagonal (hopping) and diagonal (onsite) modulations. Using the inverse participation ratio and the wave-packet centroid, we construct localization–delocalization phase diagrams for both equilibrium and nonequilibrium steady states. We analyze the robustness of initial-state properties under dissipation and characterize dissipation-induced localization–delocalization transitions (and their reversals) in detail. Trace-distance dynamics provide evidence for a quantum Mpemba effect: states prepared farther from the steady state can relax faster than states initialized closer to it. We propose a starting-line hypothesis that explains the presence or absence of this effect across parameter regimes. These results advance the understanding of steady-state phase transitions and relaxation dynamics in dissipatively driven quasiperiodic systems.

\end{abstract}

\maketitle

\section{Introduction}
Anderson localization is a pivotal quantum phenomenon in condensed-matter physics,
profoundly impacting our understanding of electron transport in disordered or
quasiperiodic systems \cite{ref1,ref2,ref3,ref4}. In three-dimensional systems,
a metal-insulator transition arises based on scaling theory \cite{ref4_1}, distinguishing
a delocalized phase (with freely propagating electrons)
from a localized phase (with spatially confined electron motion).
Anderson localization research is vital for understanding particle behavior
in complex disordered settings and advancing optical lattice experimental
design and measurement techniques\cite{ref5}.
Experimentally, Anderson localization has been observed in various platforms. For instance,  in ultra-cold atomic systems,
where precise control of interatomic interactions and external potentials
enables the creation of tunable disorder and clear observation of the
delocalization-to-localization transition \cite{ref5,ref6,ref7,ref8,ref9,ref10,ref11,ref12,ref12_1,ref12_2,ref12_3}.
Moreover, photonic quasicrystals also exhibit Anderson localization. Studying Anderson localization
in photonic crystals helps in grasping light transmission and fostering the development of functional
optical devices \cite{ref13,ref14,ref15,ref16,ref17,ref18,ref19,ref19_1}.

Under different physical mechanisms, Anderson localization takes various forms.
In quasiperiodic systems, when the quasiperiodic potential strength
exceeds a critical value, all quantum states become localized \cite{ref2,ref6}. In
one-dimensional quasiperiodic systems with short- (long-) range hoppings \cite{ref20,ref21,ref22,ref23,ref24,ref25,ref26,ref26_1}
and generalized quasiperiodic modulations \cite{ref27,ref28,ref31,ref33,ref34,ref35_4,ref35_6,ref35_7}    % ,ref29,ref30,ref32,ref35
, Anderson localization
occurs only as specific energy levels, separated from delocalized level by mobility
edges. This gives the system an intermediate-phase, neither fully delocalized nor
fully localized. Recently, the hidden self duality in the system with quasiperiodic
modulations is discovered \cite{ref35_8}, which further advances the understanding about the
Anderson localization and mobility edges \cite{ref35_9,ref35_10}.

As a powerful framework for addressing dissipation \cite{10.1093/acprof:oso/9780199213900.001.0001,ashidaNonHermitianPhysics2020a}, the burgeoning field of accurate manipulation of quantum coherence in
experiments \cite{exp_NH1,exp_NH2,exp_NH3,exp_NH4} have spurred intense interest in dissipative open quantum systems.
Dissipation proves pivotal in inducing both localized and delocalized states,
offering a crucial mechanism to decipher electron transport in disordered and
homogeneous materials \cite{dis_1,dis_2,dis_3,Longhi,Wang}    %,Wang_2
. It can disrupt localization and boost transport \cite{dis_1,dis_2,dis_3},
induce mobility edges in systems that previously lacked delocalized-localized
transitions \cite{Longhi}, and drive quasiperiodic systems into specific delocalized
or localized states \cite{Wang,Wang_3}%PRB_wang,
, while also modulating the topological properties of
topologically trivial and nontrivial insulators \cite{feng}    %,Wang_2
, and the initial property
dependence of the steady state has also been discovered \cite{feng}. This motivate us to
study the delocalization to localization transformation of the steady state of
systems with diagonal and off-diagonal quasi-periodic modulation under dissipation,
and discuss the connection between the localization properties of the steady state
and the initial properties of the system.

In studies of dissipation-induced delocalization–localization transitions, prior work has largely concentrated on the
properties of nonequilibrium steady states. Indeed, the possibility that states initialized farther from the steady (equilibrium) state may relax faster than those prepared closer to it---an instance of the quantum Mpemba effect, the quantum counterpart of the classical phenomenon---has also attracted considerable attention \cite{MP1,MP2,MP3,MP4,MP5}.
Regarding to the quantum Mpemba effect, it has been theoretically predicted across a diverse range of
systems, including quantum dots \cite{MP_dot1,MP_dot2,MP_dot3,MP_dot4}, spin or bosonic systems \cite{MP_SB_2,MP_SB_4,MP_SB_5,MP_SB_6}, %MP_SB_1,MP_SB_3,
multi-level systems \cite{MP_Ml_1,MP_Ml_2,MP_Ml_3,MP_Ml_4}, as well as other platforms \cite{MP_others_2,MP_other_PRA,Jiang_1,Jiang_2,Ares_4}.    %MP_others_1,MP_others_4,
A theoretical framework for the origin of the Mpemba effect in closed many-body quantum systems, together with concise reviews and perspectives on its quantum variants, has been presented \cite{Ares_2,Ares_3}. Experimental observations have been reported in specific platforms, including trapped-ion quantum simulators \cite{Ares_1}, consistent with theoretical predictions, and single trapped-ion systems \cite{MP_exp_2,MP_exp_3}.
Motivated by recent advances, we examine whether the quantum Mpemba effect can occur in an open, quasiperiodic system
with simultaneous diagonal and off-diagonal modulations. While earlier work established the effect for localized initial
conditions \cite{MP_other_PRA}, here we consider nonlocalized preparations and delineate the conditions under which
the effect, if any, survives. We further develop a unified framework that explains both the occurrence and the breakdown of
the quantum Mpemba effect.

This work is organized as follows. Section \ref{S2} studies the model with
diagonal and off-diagonal modulations and gives out the localization phase
diagram. Section \ref{S4} studies the delocalization-localization transition of the steady
state after considering the bond dissipations. Section \ref{S5} reveals the
phenomenon of quantum Mpemba effect and puts forward a hypothesis to explain
the cause of this phenomenon. A summary is presented in Sec.~\ref{S6}.

\begin{figure}[t]
		\centering
		\includegraphics[width=0.5\textwidth]{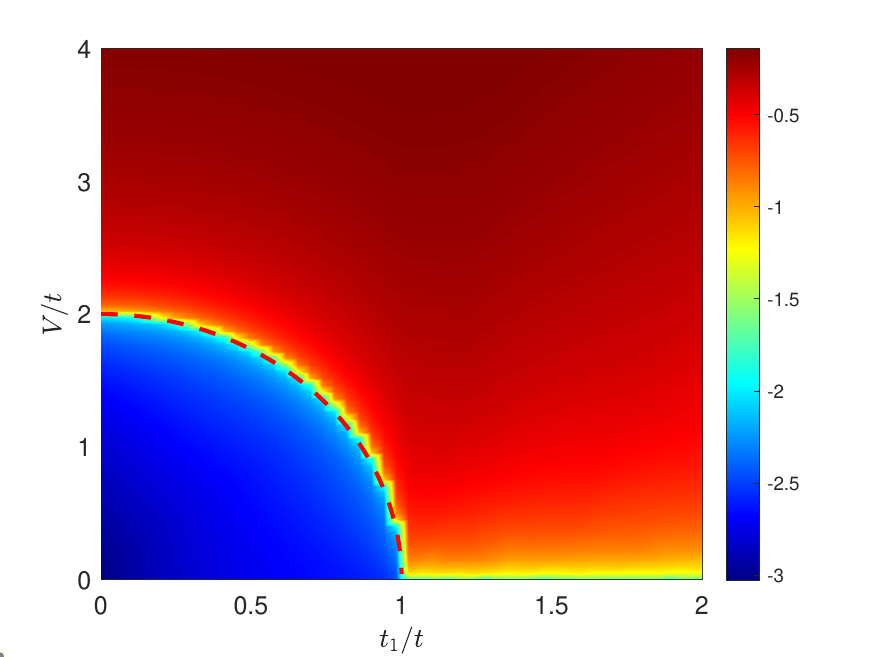}\\
		\caption{(Color Online) The phase diagram illustrates how $\log_{10}(\rm MIPR)$
varies with $V/t$ and $t_{1}/t$, where the system size is set to $L = 1597$.
The red dashed line represents the phase boundary, which serves to distinguish the
extended phase from the localized and critical phases.  Meanwhile, the colorbar
corresponds to the numerical values of $\log_{10}(\rm MIPR)$.}
\label{f1}
\end{figure}

\section{Model and localization phase diagram}\label{S2}
We study the quasiperiodic system with off-diagonal and diagonal quasiperiodic modulation,
whose Hamiltonian is presented as
\begin{equation}
H=\sum_{n}\left(t_{n}\hat{c}^{\dag}_{n}\hat{c}_{n+1}+H.c.\right)+\sum_{n}V_{n}\hat{c}^{\dag}_{n}\hat{c}_{n},
\end{equation}
where $t_{n}=t+t_{1}\cos{(2\pi\alpha n)}$ and $V_{n}=V\cos{(2\pi\alpha n)}$. $t$ is chosen as the unit of energy, $n$ is the site index, and $\alpha=\frac{\sqrt{5}-1}{2}$ induces
the incommensurate quasiperiodic modulations.

\begin{figure}[t]
		\centering
		\includegraphics[width=0.5\textwidth]{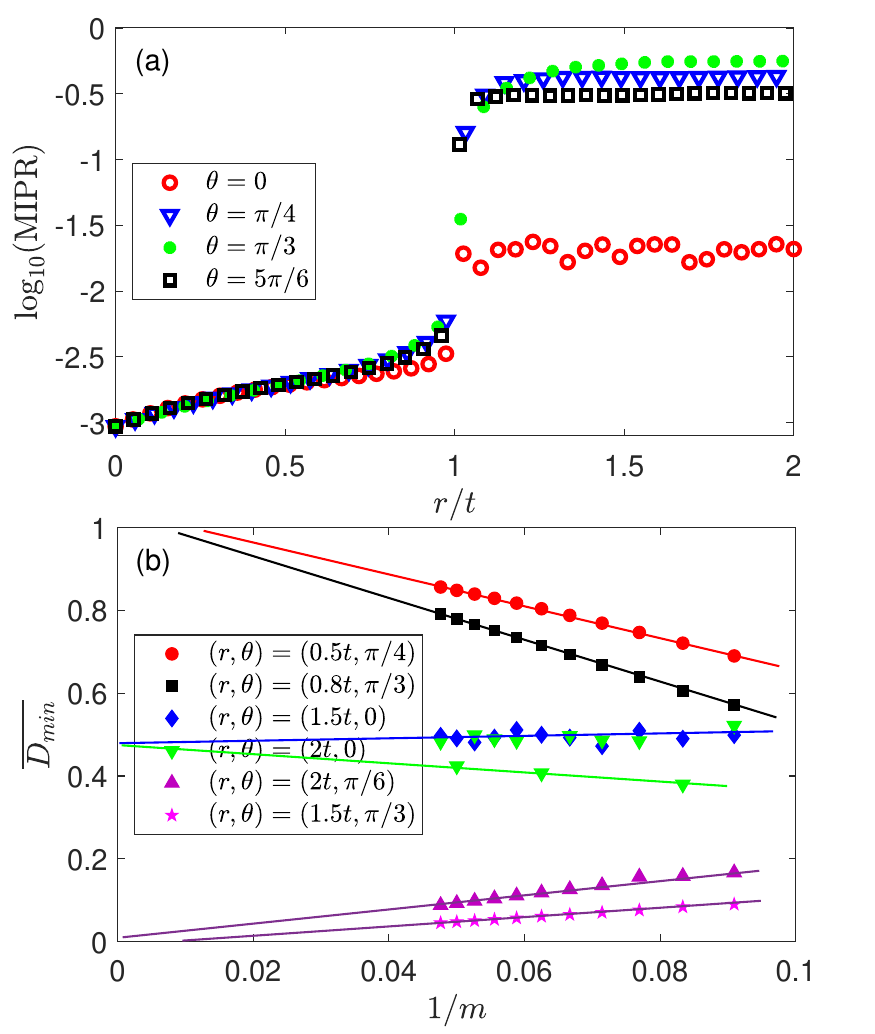}\\
		\caption{(Color Online) (a) $\log_{10}(\rm MIPR)$ as a function of $r$ with
$\theta=0$, $\pi/4$, $\pi/3$, and $5\pi/6$.
(b) $\overline{D_{min}}$ as a function of the inverse Fibonacci index $1/m$ under
different $(r,\theta)$.
}
		\label{f2}
\end{figure}

The localization phase diagram can be determined from the inverse participation ratio (IPR).
Given a normalized wave function $\vert\psi_{j}\rangle=\sum^{L}_{n=1}\phi_{j}(n)\hat{c}^{\dag}_{n}|0\rangle$ (Here $j$ is the index of wave function),
the corresponding ${\rm IPR}_{j}$ is
\begin{equation}
{\rm IPR}_{j}=\sum^{L}_{n=1}|\phi_{j}(n)|^{4}.
\end{equation}
It is known that the ${\rm IPR}$ of the extended, critical, and localized state,
respectively, possess the properties of ${\rm IPR}\approx 0$, $0<{\rm IPR}<1$, and ${\rm IPR} \approx 1$ \cite{ref20}.
Under the given parameters, the extended, critical, and localized phase
can be further characterized by the average value of ${\rm IPR}$ of all wave functions,
namely the mean inverse participation ratio (MIPR), denoted
as ${\rm MIPR}=\sum^{L}_{j=1}{\rm IPR}_{j}/L$. With system size $L=1597$,
the localization phase diagram in $V$-$t_{1}$ parameter space is plotted in Fig.~\ref{f1},
where the color bar denotes $\log_{10}(\rm MIPR)$.
In fact, for the $V=0$ case, the earlier studies have proved that the
system exhibits an extended-critical phase transition \cite{Tong_Gao}. We can see from Fig.~\ref{f1}(a)
that under $V=0$, $\log_{10}(\rm MIPR)$ for the extended phase approaching $-3$ and
$\log_{10}(\rm MIPR)$ for the critical phase approaches \textcolor{red}{$-1.6$}. Accordingly, for finite $V$, the
quarter-elliptic region in parameter space where $\log_{10}(\rm MIPR) \approx -3$ is identified with
the extended (delocalized) phase, whereas outside this region $\log_{10}(\rm MIPR)>-1$ indicates localization.

To locate the phase boundary of the extended-localized transition, we shall introduce the
transformation $r=\sqrt{V^{2}/4+t^{2}_{1}}$ and
$\theta={\rm Arctan}(V/t_{1})$ ($\in \left[0,2\pi\right]$). Under $L=1597$,
the $\log_{10}(\rm MIPR)$ versus $r$ with different $\theta$ are plotted
in Fig.~\ref{f2}(a). Intuitively, we can see that for different $\theta$,
the $\log_{10}(\rm MIPR)$-$r$ curves characterizing the extended-localized (critical)
transition all present a jump at the critical value $r_{c}/t=1$, namely the phase
boundary $\sqrt{V^{2}/4+t^{2}_{1}}=t$ extracted from the numerical results, which has been plotted as
the red dashed line in Fig.~\ref{f1}.

To further verify the above mentioned conclusions, we introduce
the fraction dimension $D$. Consider a system where the system size
$L$ equals the $m$-th Fibonacci number $F_{m}$ and the incommensurate parameter $\alpha$
is replaced by $\alpha=F_{m-1}/F_{m}$. At a specific lattice site $n$,
the fraction dimension $D_{n}$ can be derived from the formula
\begin{equation}
p_{n}=F_{m}^{-D_{n}}.
\end{equation}
Here, $p_{n}$ represents the probability density. This equation reveals that
$D_{n}$ acts as a scaling index. For an extended state, since $p_{n} \sim 1/F_{m}$,
we have $D_{n}\sim 1$. In the case of a localized state, $D_{n}\sim 1$ at the sites
where the particle occupies, while $D_{n}\rightarrow \infty$ at unoccupied sites.
For a critical state, $D_{n}$ lies within a finite interval $\left[D_{min},D_{max}\right]$.
Thus, the minimum value $D_{min}$ directly reflects the nature of a given wave function
$\vert\psi_{j}\rangle$. Specifically, $D_{min} \rightarrow 0$ indicates a localized state,
$0< D_{min}<1$ points to a critical state, and $D_{min}\rightarrow 1$ signifies
an extended state. For generality, we use the average of $D_{min}$ over all states,
defined as $\overline{D_{min}}=\sum^{L}_{j=1}D^{j}_{min}/L$. In the limit of
$1/m \rightarrow 0$ (extrapolation limit), this average helps distinguish different phases.
We select representative parameter points in various phases to compute $\overline{D_{min}}$.
As depicted in Fig.~\ref{f2}(b), we observe that the relevant $D_{min}$ approaches
$1$ at the parameter sites $\left(r,\theta\right)=\left(0.5t,\frac{\pi}{4}\right)$
and $\left(r,\theta\right)=\left(0.8t,\frac{\pi}{3}\right)$. These observations
confirm that the system resides in the extended phase under these conditions.
As anticipated, the corresponding $D_{min}$ values fall within the
interval $\left(0,1\right)$ in the thermodynamic limit when
$\left(r,\theta\right)=\left(1.5t,0\right)$ and $\left(r,\theta\right)=\left(2t,0\right)$.
These results clearly reveal the critical-phase characteristics of the system in such
parameter regimes. When $\left(r,\theta\right)=\left(1.5t,\frac{\pi}{3}\right)$ and
$\left(r,\theta\right)=\left(2t,\frac{\pi}{6}\right)$, the associated $D_{min}$
extrapolates to $0$. This outcome identifies that the system is in the localized
phase at these parameter points.

\section{Dissipation driven delocalization-localization transition}\label{S4}
Recent research efforts have yielded
conflicting perspectives regarding the steady-state delocalization-localization
transition under dissipative modulation \cite{Longhi,Wang,PRB_wang}    %,Wang_3
. By employing the dissipation,
one can precisely control the delocalization and localization properties of
the system under non-equilibrium steady states. Therefore, it motivates us to
study the delocalization and localization properties of the steady state after
introducing the dissipation. Besides, some studies suggest this transition is independent of the system's
initial properties \cite{Wang}. Alternative viewpoints argue that the system's
initial attributes can influence the transition. Different initial property may either shorten
or extend the threshold required for the delocalization-localization shift to occur.
We aim to study whether the dependence of system's initial property will also occur in
this system with diagonal and off-diagonal quasiperiodic modulations and to study
how different initial properties affect the delocalization-localization transition
of the steady state as well.

Upon introducing this bond dissipation Lindblad operator $L_{n}=c^{\dag}_{n}c_{n+\ell}$,
generated following the ones in Ref.\cite{bond_1,bond_2,bond_5,bond_6,Wang}    %bond_3,bond_4,
, then the time-evolution of the density
matrix $\rho$ is dictated by the following Lindblad master equation
\begin{equation}
\dot{\rho}=-\left[H,\rho\right] + \sum_{n}\gamma_{n}\mathcal{D}\left[L_{n}\right]\rho,
\end{equation}
Here, the dissipator $\mathcal{D}\left[L_{n}\right]$ is defined as
$\mathcal{D}\left[L_{n}\right]=L_{n}\rho L^{\dag}_{n}-\{L^{\dag}_{n}L_{n},\rho\}/2$,
and $\gamma_{n}=\gamma\cos({2\pi\alpha n})$ represents the tunable quasiperiodic bond
dissipation strength with the incommensurate parameter $\alpha=\frac{\sqrt{5}-1}{2}$.
The rationale for our selection of the quasiperiodic form for the key dissipation
operator stems from the consideration of the following aspect: the initial key dissipation
operator is made up of four terms, with two being dephasing terms \cite{Longhi}
and the remaining two being hopping terms. As indicated in Ref.~\cite{Wang}, the dephasing effect fails to
induce the delocalization-localization transition of steady states. It has been theoretically
and experimentally confirmed that quasiperiodic modulation is capable of causing
localization \cite{ref6}. Moreover, this bond dissipation operator can be experimentally realized by introducing the auxiliary lattice \cite{auxiliary_1,auxiliary_3,auxiliary_4}. 
Owing to these factors, we utilized such a dissipative operator.
This evolution equation can also be equivalently expressed as
\begin{equation}
\rho(\tau)=e^{\mathcal{L}\tau}\rho(0),
\end{equation}
which encapsulates all the system's dynamical information throughout the
evolution process. Given that the real components of the eigenvalues of
the Lindbladian matrix $\mathcal{L}$ are non-positive, as time evolves
to the long-time limit, the density matrix will ultimately relax to the
zero-energy eigenstate of $\mathcal{L}$, namely the non-equilibrium
steady state (NESS) $\rho_{ss}$.

\begin{figure}[htp]
		\centering
		\includegraphics[width=0.5\textwidth]{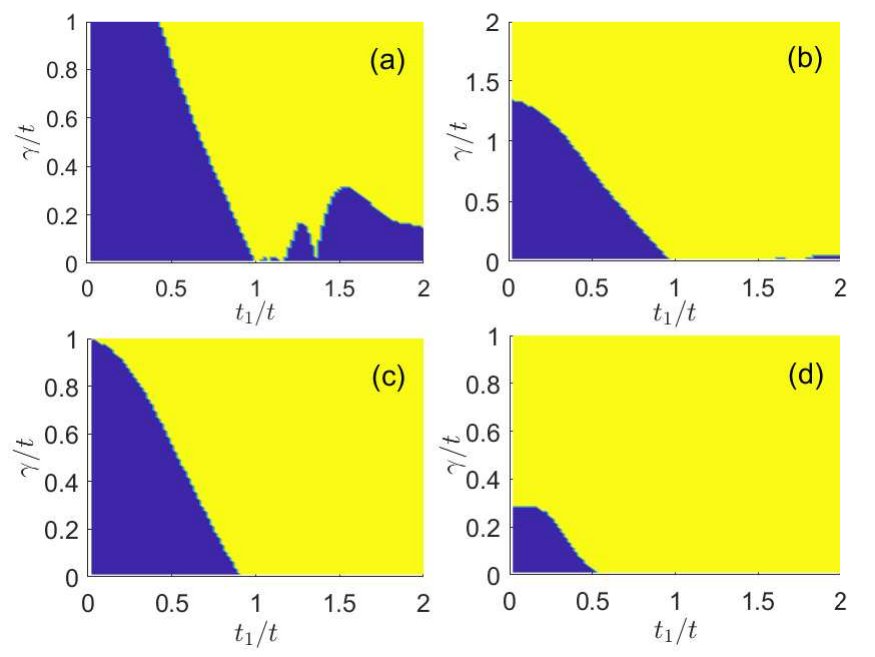}\\
		\caption{(Color Online) Steady state phase diagram.
(a) $V=0$. (b) $V=0.5t$.
(c) $V=t$. (d) $V=2t$.
The blue regions denote the delocalization phases. The yellow
regions denote the localization phases.
Other parameter is $L=100$.
}
		\label{f5}
\end{figure}

With $L=100$ and $\ell=1$, the delocalization phase diagrams for
$V=0$, $V=0.5t$, $V=t$, and $V=2t$ are plotted in Figs.~\ref{f5}(a)-
\ref{f5}(d), respectively. The phase diagram is obtained by analyzing
the position of the centroid in a steady state. We define a steady state
where the centroid position (its definition is given in Eq.(\ref{x_cm_ss})) is no more than five times the lattice constant
from the system boundary as a localized state; otherwise, it is a delocalized
state. Our choice of criterion is guided by the experiment reported in Ref. \cite{ref6},
which realized a one-dimensional Aubry–André (AA) model in a cold-atom optical lattice and observed
the Anderson-localization transition; in the localized regime, the wave packet of an initially localized
state broadened by approximately 10 lattice constants.
Therefore, it is feasible to distinguish the delocalization-localization transition of the
steady state through the position of the center of mass.

As can be seen from Figs.~\ref{f5}(a)-\ref{f5}(c), when $t_{1}$ is less
than $0.5t$, there is a large parameter range that makes all the steady
states delocalized. This indicates that under these parameters, the steady
state can retain the initial delocalized property of the system.
As $\gamma$ or $t_{1}$ increases, the steady state gradually changes
from a delocalized state to a localized state, meaning that the properties
of the steady state can be regulated by tuning $\gamma$ and $t_{1}$.
Meanwhile, we observe that as $V$ gradually increases,
the range of the delocalized region gradually shrinks. It is indicated that
adjusting the parameter $V$ can also regulate the localization property of
the steady state. For instance, when V is at a relatively small value,
it can be seen from \ref{f5}(a) and \ref{f5}(b) that when $t_{1}>t$, there
are still delocalized states within it, resulting in the reentrance
delocalization phenomenon. Moreover, the value of $V$ is smaller,
making the delocalization region larger. When $V$ is relatively large,
it can be seen from Figs.~\ref{f5}(c) and \ref{f5}(d) that the delocalized
region disappears. In particular, we find that when $V=2t$, even if the
initial system is in the localized phase, the steady state is still
delocalized (as can be seen from Fig.~\ref{f5}(d), the delocalized state
exists within a small range of $t_{1}$ and $\gamma$).

\section{quantum Mpemba effect and starting-line hypothesis}\label{S5}
Regarding the delocalization-localization transition in open quasiperiodic
system, the focus on this topic mainly lies in the localization properties of
steady states. For an anomalous dynamic process: the relaxation rate of a
system whose initial state is further from the equilibrium state is actually
faster than that of a system whose initial state is closer to the equilibrium
state, that is, the quantum Mpemba effect, remains to be studied and discussed.

\begin{figure}[htp]
		\centering
		\includegraphics[width=0.5\textwidth]{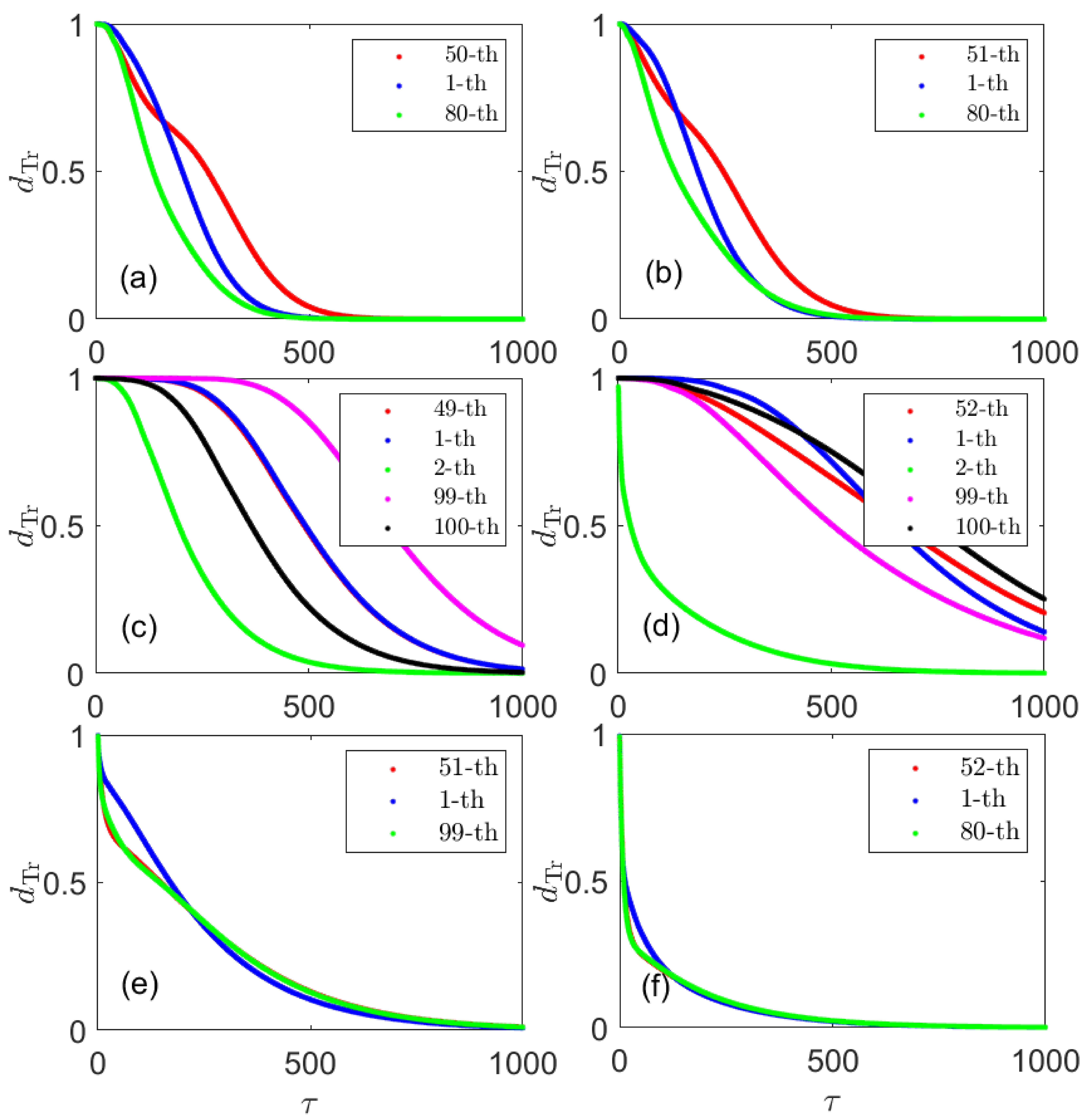}\\
		\caption{(Color Online) Trace distance $d_{\rm Tr}$ versus evolution time $\tau$.
(a) $V=0$, $t_{1}=1.2t$, and $\gamma=0.5t$. (b) $V=0$, $t_{1}=1.7t$, and $\gamma=0.8t$.
(c) $V=0.2t$, $t_{1}=1.2t$, and $\gamma=0.2t$. (d) $V=0.2t$, $t_{1}=1.5t$, and $\gamma=0.2t$.
(e) $V=0.2t$, $t_{1}=0.8t$, $\gamma=0.2t$. (f) $V=0.5t$, $t_{1}=0.5t$, $\gamma=0.2t$.
The ordinal number in the caption indicates the ordinal number of the initial state in the eigenstates of $H$. 
Other parameter is $L=100$.
}
		\label{f6}
\end{figure}
%\textcolor{red}{The blue regions denote the delocalization phases. The yellow regions denote the localization phases.}
%Specifically, the initial state energy corresponding to the red curve is closer to the steady-state energy than that of other initial states.

We employ the density matrix $\rho(\tau)$ that describes the evolution of the
system over time and the steady state density matrix $\rho_{ss}$ to
calculate the standard distance between the matrices, this is, the trace
distance, to characterize the quantum Mpemba effect. The trace distance
$d_{\rm Tr}$ is defined as
\begin{equation}
d_{\rm Tr}=\frac{1}{2}{\rm Tr}\sqrt{T^{\dag}T},
\end{equation}
where $T=\rho(\tau)-\rho_{ss}$.
Under Lindbladian dynamics, this quantity is monotonic in time and has been employed experimentally to probe both the strong quantum Mpemba effect and its inverse counterpart \cite{MP_exp_2,MP_exp_3}. A nonzero trace distance $d_{\rm{Tr}}>0$ indicates that the system has not yet relaxed to the steady state, whereas $d_{\rm{Tr}}=0$ certifies convergence to the steady state.

Before examining the quantum Mpemba effect, we introduce the starting-line hypothesis.
Consider a set of initial states characterized by their energies and center-of-mass (COM) positions.
Let $E_{\mathrm{ss}}$ and $x_{\mathrm{com}}^{\mathrm{ss}}$ denote the energy and COM of the steady state,
and define the COM distance as
\begin{eqnarray}
	\Delta x_{\mathrm{com}} \equiv\left|x_{\mathrm{com}}(0)-x_{\mathrm{com}}^{\mathrm{ss}}\right|,
\end{eqnarray}
where the center-of-mass position of $\ket{\psi(0)}$ at the moment $t=0$ (the initial state, selected 
from the $j$-th eigenvector of $H$) has been define as
\begin{eqnarray}
	x_{\mathrm{com}}(0)=\sum_{n=1}^{L}\left(|\bra{n}\ket{\psi(0)}|^2 \times n \right)
\end{eqnarray}
and for the steady-state $\ket{\psi^{ss}}$ achieved after sufficient evolution, its center-of-mass position is defined as
\begin{eqnarray}
	x_{\mathrm{com}}^{ss}=\sum_{n=1}^{L}\left(|\bra{n}\ket{\psi^{ss}}|^2 \times n \right)
	\label{x_cm_ss}
\end{eqnarray}
where $n$s represent as the marks of the site index.
The hypothesis posits: if the state whose energy is close to $E_{\mathrm{ss}}$ has a larger $\Delta x_{\mathrm{com}}$ than
another state whose energy lies farther from $E_{\mathrm{ss}}$, then the latter can relax faster, manifesting the quantum Mpemba effect.
This yields a simple criterion for which initial states can trigger the effect. For extended states, $x_{\mathrm{com}}(0)$
is pinned to the system center, so candidate extended initial states share almost the same 'starting line' and $\Delta x_{\mathrm{com}}$
do not exhibit the effect. By contrast, localized and critical states generally have $x_{\mathrm{com}}(0)$ displaced
from the center; such 'starting-line discrepancies' enable a state that is farther in energy (but closer in COM) to relax
faster than a state that is closer in energy (but farther in COM), thereby producing the quantum Mpemba effect.

We take three different types of system eigenstates as the initial
states and calculated the evolution of trace distance over time, as
shown in Figures \ref{f6}(a)-\ref{f6}(f). The ordinal number in the caption indicates
the ordinal number of the initial state in the eigenstates of $H$.
Specifically, the initial state energy corresponding to the red curve is closer to the
steady-state energy than that of other initial states. Figures \ref{f6}(a) and \ref{f6}(b) show the results of the delocalized states with critical properties as
the initial state. The COM positions corresponding to the initial
states (rounded off, the same below) are respectively: $n=\{60,49,39\}$
and $n=\{60,49,41\}$, and the COM distances are respectively:  $\Delta x_{\mathrm{com}}=\{57, 47, 36\}$ and
$\Delta x_{\mathrm{com}}=\{56, 46, 38\}$. It can be seen from the data that the COM distance $\Delta x_{\mathrm{com}}$ between the initial state
with energy close to the steady state and the steady state is actually greater than that when the energy is far from
the steady state. The time required for $d_{\rm Tr}$ to evolve to zero is longer, and thus the so-called quantum
Mpemba effect emerges.

Figures \ref{f6}(c) and \ref{f6}(d) show the
results of the localized states as the initial states, where the corresponding COM
positions of the initial states are respectively: $n=\{63,61,30,83,49\}$
and $n=\{66,63,15,56,71\}$, and the COM distances are respectively: $\Delta x_{\mathrm{com}}=\{60, 58, 27, 80, 46\}$ and
$\Delta x_{\mathrm{com}}=\{61, 58, 10, 51, 66\}$. We can observe that when energy moves away from the initial state of
steady-state energy, it evolves to the steady state more quickly compared to the quantum state where energy is close
to the initial state energy. Therefore, the so-called quantum Mpemba effect also emerges.
We can still observe  that for the initial state where energy is far from the steady-state energy,
it evolves to the steady state more quickly than the quantum state where energy is close to the steady-state energy.
Therefore, the so-called quantum Mpemba effect emerges.  This phenomenon can  be explained by the proposed
hypothesis as well, because the COM of the initial state with energy close to the steady state is further away from
the steady state compared to other initial states, and thus it takes longer time to evolve to the steady state.

Figures \ref{f6}(e) and \ref{f6}(f) present
delocalized states with extended properties as the results of initial
states, where the corresponding COM positions of the initial states
are respectively: all $n=50$ and all $n=51$, and the COM distances are
respectively: $\Delta x_{\mathrm{com}}=42$ and
$\Delta x_{\mathrm{com}}=21$. We can see that regardless of whether the energy is close to or far
from the initial state of the steady-state energy, the time it takes for $d_{\rm Tr}$ to evolve to zero is
almost indistinguishable, and thus the so-called quantum Mpemba effect does not occur. This phenomenon
can be explained by using our hypothesis. Because the initial COM positions and COM distances of different
initial states are almost the same, the time required for evolution to a steady state is also almost the same.
Based on these results, we can conclude that for the
initial states with extended properties, the dynamical evolution behavior
does not exhibit the quantum Mpemba effect, while for the initial states
with critical and localized properties, the dynamical evolution behavior will
exhibit the quantum Mpemba effect. From the perspective of the COM
position of the initial state, when the Mpemba effect occurs, the wave
packet of the initial state is relatively closer to the boundary of the system,
which confirms the starting-line hypothesis we proposed.

\section{Summary}\label{S6}

We have investigated a one-dimensional tight-binding model with simultaneous quasiperiodic modulations
of both hopping and onsite potential, focusing on localization–delocalization transitions under
equilibrium and nonequilibrium conditions, and quantum relaxation dynamics. Using finite-size numerics, we have mapped
the equilibrium phase diagram and verified it via the mean inverse participation ratio and fractal dimensions,
finding that the delocalized (extended) and localized phases are separated by an approximately elliptical phase boundary. 
By computing the nonequilibrium steady-state phase diagram under quasiperiodic bond dissipation and
analyzing initial-state dependence, we have known: for moderate $V$ and $t_{1}$, the steady state largely inherits the
initial state’s delocalized character; tuning $\gamma$, $t_{1}$ or $V$ continuously drives the steady state from delocalized
to localized, and for sufficiently small $V$ the steady state can become delocalized even from localized initial conditions.
Finally, using the trace distance, we have demonstrated that there exists quantum Mpemba effect in which
states initialized in localized or critical regimes can relax faster than states prepared closer to the steady state.
These observations are rationalized by a starting-line hypothesis, whose applicability to broader classes of
open quasiperiodic systems remains an open question.

This research is supported by Zhejiang Provincial Natural Science Foundation of China under Grant No. LQN25A040012,
the National Natural Science Foundation of China under Grant No. 12174346, and the start-up fund from Xingzhi College, Zhejiang
Normal University.

\bibliography{references}

\end{document}